\documentclass[11pt]{article}

\usepackage[final]{acl}

\usepackage{times}
\usepackage{helvet} % ensure Helvetica (sans) is embedded when used by figures
\usepackage{latexsym}

\usepackage[T1]{fontenc}
\usepackage[utf8]{inputenc}

\usepackage{microtype}

\usepackage{inconsolata}

\usepackage{graphicx}
\usepackage{subcaption}

\usepackage{amsmath,amssymb,amsfonts}
\usepackage{algorithmic}
\usepackage{graphicx}
\usepackage{textcomp}
\usepackage{xcolor}
\usepackage{url}

\usepackage{listings}
\lstdefinelanguage{JavaScript}{
  keywords={break, case, catch, continue, debugger, default, delete, do, else, finally, for, function, if, in, instanceof, new, return, switch, this, throw, try, typeof, var, void, while, with, let, const},
  keywordstyle=\color{blue}\bfseries,
  ndkeywords={class, export, boolean, throw, implements, import, this},
  ndkeywordstyle=\color{darkgray}\bfseries,
  identifierstyle=\color{black},
  sensitive=false,
  comment=[l]{//},
  morecomment=[s]{/*}{*/},
  commentstyle=\color{gray}\ttfamily,
  stringstyle=\color{red}\ttfamily,
  morestring=[b]',
  morestring=[b]"
}
\usepackage{color}
\usepackage{xspace}
\usepackage{algorithm}
\usepackage{booktabs}
\usepackage{xltabular}
\usepackage{wrapfig}
\usepackage{float}      % for fixing the position of minipage table
\usepackage{caption}    % for adding captions to minipages

\usepackage{hyperref}   % Ensure hyperref is loaded before hyperxmp
\usepackage{hyperxmp}
\usepackage{microtype}
\usepackage{multirow}
\usepackage{placeins}
\usepackage{needspace}

\usepackage{tcolorbox}

\newcommand{\TheName}{\textsc{SynthFix}}

\def\BibTeX{{\rm B\kern-.05em{\sc i\kern-.025em b}\kern-.08em
    T\kern-.1667em\lower.7ex\hbox{E}\kern-.125emX}}

\begin{document}

% \title{\TheName{}: Neuro-Compiler Vulnerability Repair via Programmer-Inspired Adaptation}

\title{\TheName{}: Adaptive Neuro-Symbolic Code Vulnerability Repair}

% \author{Anonymous Author(s)}

\author{
    \textbf{Yifan Zhang}\textsuperscript{1}\hspace{5pt}
    \textbf{Jieyu Li}\textsuperscript{1}\hspace{5pt}
    \textbf{Kexin Pei}\textsuperscript{2}\hspace{5pt}
    \textbf{Yu Huang}\textsuperscript{1}\hspace{5pt}
    \textbf{Kevin Leach}\textsuperscript{1}\vspace{5pt}\\
    % \textbf{Jiahao Zhang}\textsuperscript{1}\vspace{5pt}\\
    Vanderbilt University\textsuperscript{1}\hspace{5pt}
    University of Chicago\textsuperscript{2}\hspace{5pt}
    \\
    \normalsize
    \texttt{\{yifan.zhang.2,yu.huang,kevin.leach\}@vanderbilt.edu}\\
    \normalsize
    \texttt{kpei@cs.uchicago.edu}\hspace{5pt}
}

\maketitle

% \begin{abstract}
% Recent advances in Large Language Models (LLMs) have shown promise for automated code repair, yet these models often struggle with complex semantic errors. We present \TheName{}, a hybrid Neuro-symbolic framework for LLM-based vulnerability repair that unifies Neuro code synthesis with compiler-informed feedback. Specifically, our method comprises: (1) a Repair Agent leveraging either Supervised Fine-Tuning~(SFT) or Reward Fine-Tuning (RFT) in fine-tuning, (2) an offline reward model that integrates symbolic signals from AST, CFG, and vulnerability detection tools to assess patch quality, and (3) a Neuro Router Model that adaptively directs each training batch to the optimal strategy based on real-time code features. Compared to the best baseline using either SFT or RFT alone, \TheName{} achieves up to 18\% relative improvement in CodeBLEU/CrystalBLEU and 32\% in Exact Match on the FixJS (JavaScript) and CodeFlaws (C) benchmarks. Our results show that adaptively combining static fine-tuning and dynamic reward-based refinement, mirroring how programmers alternate between pattern fixes and compiler feedback, considerably improves LLM vulnerability repair by boosting accuracy and convergence. Our code and data are available at \url{https://github.com/CoderDoge1108/SynthFix}.

% \end{abstract}

\begin{abstract}
Large Language Models (LLMs) can generate plausible code patches, but plausibility is not enough for automated repair: a patch must compile, pass tests, and remove the target vulnerability. We present \TheName{}, a neuro-symbolic repair framework that combines supervised repair learning with compiler-informed feedback. During training, a lightweight router selects between Supervised Fine-Tuning (SFT) for common repair patterns and Reward Fine-Tuning (RFT) for examples that benefit from symbolic feedback. The reward combines static structure, lint/compile checks, security scanning, and public execution tests where available; at inference time, the same evidence guides best-of-$K$ candidate selection under a greedy floor. Across five code LLMs (1.3B--7B) on pyrepair, CodeFlaws, and SVEN, \TheName{} improves deployable repair metrics over SFT-only and RFT-only baselines, with relative gains up to 54\% in functional correctness and 14\% in security clearance. Our code and data are available at \url{https://github.com/CoderDoge1108/SynthFix}\footnote{Our artifacts are publicly available under the MIT License.}.
\end{abstract}

\begin{figure*}[t]
\centering
\includegraphics[width=1\linewidth]{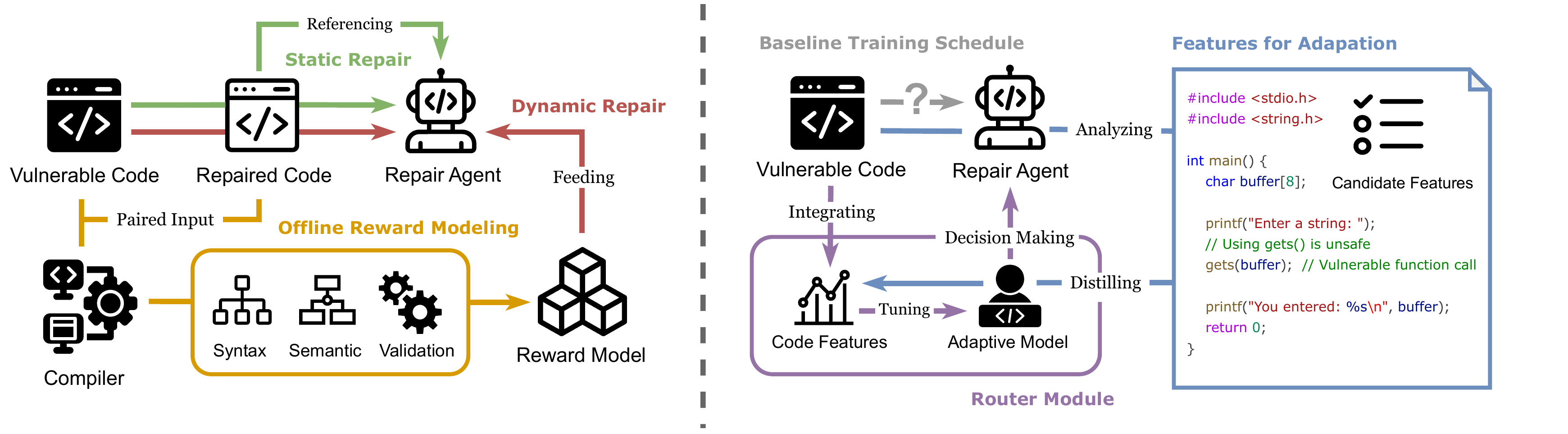}
\caption{Overview of \TheName{}. The repair agent generates candidate patches, the symbolic reward model provides compiler-, analyzer-, and test-derived feedback, and the router adaptively chooses between SFT and RFT during training.}
\label{fig:SynthFix_main_draft}
\end{figure*}

\section{Introduction}

Large Language Models (LLMs) can generate and edit source code with impressive fluency~\cite{chen2021evaluating, allamanis2018survey}. For automated vulnerability repair, however, fluent code is not enough: a patch must compile, preserve intended behavior, pass available tests, and remove the target vulnerability~\cite{fu2022vulrepair}. Neural models often optimize for textual plausibility and local syntactic patterns, yielding patches that look reasonable but fail functionally or leave the vulnerability intact~\cite{berabi2021tfix, jiang2023impact}. This motivates repair systems that combine neural generation with compiler- and analyzer-derived evidence~\cite{shi2020pathpair2vec, wu2022turn}.

The standard adaptation strategy is Supervised Fine-Tuning (SFT) on bug-fix pairs~\cite{dong2023abilities}. SFT is efficient for common repair idioms, but it provides limited pressure toward executable or security-cleared behavior beyond the demonstrations. Reward Fine-Tuning (RFT) offers a complementary path: models can optimize against public tests, security scans, or other task feedback~\cite{le2022coderl,islam2024code,fang2025dpo,shojaee2023execution}. In practice, applying RFT uniformly can be costly and unstable, and our experiments show that a fixed RL schedule alone does not reliably improve deployable repairs over a strong SFT baseline.

We present \TheName{}, an adaptive neuro-symbolic repair framework that improves on a fixed fine-tuning recipe in two ways. First, it treats SFT and RFT as complementary repair modes: common edits are learned through direct imitation, while harder examples receive reward-driven updates. Second, it keeps symbolic evidence active at inference time, where sampled candidates are ranked by compiler-, analyzer-, and test-derived signals under a greedy floor. Symbolic feedback therefore serves both as a training signal and as the decision rule for selecting the final patch.

The framework constructs its procedural reward from complementary signals, including AST/CFG/DFG structure, linting and compile checks, security scanning, and public execution tests when available. These signals connect neural code generation to the kinds of evidence used by program-analysis tools and neuro-symbolic repair systems~\cite{parisotto2016neuro}. A lightweight neural router then controls training. Rather than committing every batch to the same objective~\cite{shen2020survey, de2024enhanced}, the router selects the SFT or RFT pathway using compiler-derived features such as code length, AST complexity, and CFG depth. This design keeps the fluent SFT prior as a stabilizing anchor while exposing harder examples to reward-driven refinement~\cite{kulsum2024case}. At inference time, \TheName{} samples a small candidate set and applies symbolic best-of-$K$ selection with a greedy floor, ensuring that exploration does not regress below the greedy patch.

We evaluate \TheName{} on three execution- and security-grounded benchmarks: pyrepair (Python, built from MBPP~\cite{austin2021mbpp} and QuixBugs~\cite{lin2017quixbugs}), CodeFlaws for C~\cite{tan2017codeflaws}, and SVEN for security repair~\cite{he2023sven}. Across five modern code LLMs (1.3B--7B), including DeepSeek-Coder~\cite{guo2024deepseekcoder}, Llama-3.2~\cite{dubey2024llama3}, Qwen3~\cite{yang2025qwen3}, CodeLLaMA~\cite{roziere2023code}, and StarCoder2~\cite{lozhkov2024starcoder}, \TheName{} improves deployable repair metrics over SFT-only and RFT-only baselines. The gains reach 54\% relative functional improvement on CodeFlaws and 14\% relative security-clearance improvement on SVEN, and module analyses show that both router-gated training and symbolic test-time selection contribute.

% Technical background is provided in Appendix~\ref{sec:appendix_1}.

% \input{sections/section7} % From 7 to 3
\section{Approach}
\label{sec:approach}

\TheName{} is a neuro-symbolic training and inference framework for code repair. Its central idea is to use symbolic evidence twice: during training, to decide when feedback-driven RFT should replace direct SFT; and at inference time, to choose the final patch from sampled candidates. The framework trains one \textbf{Repair Agent} and adds two lightweight components: a symbolic \textbf{Reward Model} that scores patches using compiler-, analyzer-, and test-derived evidence, and a neural \textbf{Router Model} that selects SFT or RFT for each batch. Figure~\ref{fig:SynthFix_main_draft} gives the architecture.

\subsection{Problem Formulation and Notation}

Let $\mathcal{D}=\{(x_i,y_i)\}_{i=1}^{N}$ denote a repair dataset, where $x_i$ is a buggy program or function and $y_i$ is the reference repair when one is available. The Repair Agent is a conditional code model $\pi_\theta(y\mid x)$ parameterized by LoRA-adapted weights $\theta$. Given $x$, the model can produce a greedy patch $\hat{y}^{(0)}$ or a sampled set of candidate patches $\mathcal{C}_K(x)=\{\hat{y}^{(0)},\hat{y}^{(1)},\ldots,\hat{y}^{(K-1)}\}$.

For any candidate $\hat{y}$, we compute a symbolic evidence vector $\boldsymbol{s}(x,\hat{y})=[s_{\text{parse}},s_{\text{static}},s_{\text{struct}},s_{\text{sec}},s_{\text{exec}},s_{\text{repair}}]\in[0,1]^m$, whose components summarize parseability, static validity, structural similarity, security-scanner feedback, public-test execution, and repair-directed edit evidence. Held-out tests are never included in $\boldsymbol{s}$; they are used only for final evaluation. \TheName{} uses this evidence vector in two places: the reward $r(x,\hat{y})$ used for RFT updates and the selector score $q(x,\hat{y})$ used to choose the final patch at inference time.

\subsection{Symbolic Reward Model for Patch Evaluation}
\label{sec:reward_model}

The Reward Model assigns each candidate patch a procedural score rather than relying on a single binary pass/fail outcome: $r(x,\hat{y})=\boldsymbol{\lambda}^{\top}\boldsymbol{s}(x,\hat{y})=\sum_{c=1}^{m}\lambda_c s_c(x,\hat{y})$, where $\lambda_c$ controls the contribution of each symbolic component. This reward is deterministic and reference-free except for optional structural comparison terms used when a reference repair is available. It gives RFT feedback about syntax, structure, security, and executable behavior without consulting held-out tests.

The main symbolic components are:
\begin{description}
    \item[Syntactic soundness ($r_{\text{AST}}$)] rewards patches that parse into well-formed syntax trees and penalizes malformed generations~\cite{wang2021syncobert}.
    \item[Structural fidelity ($r_{\text{CFG}}$, $r_{\text{DFG}}$)] measures whether control- and data-flow structure remains consistent with the intended repair~\cite{liang2019jsac}.
    \item[Security and execution evidence ($r_{\text{Semgrep}}$, $r_{\text{exec}}$)] rewards patches that remove detected vulnerability patterns and, when public tests are available, execute successfully.
\end{description}
Together these signals guide the model toward patches that are syntactically valid, structurally plausible, security-aware, and executable. The reward is intentionally composite: no single signal is assumed to be sufficient across all benchmarks. For example, public tests are highly informative when present, while Semgrep and static validity checks remain available for security settings where executable tests may be absent.

\subsection{Adaptive Training with a Router Model}

The Router Model decides whether each batch should update the Repair Agent through SFT or RFT. For a mini-batch $B\subset\mathcal{D}$, we compute a compiler-derived routing feature vector $\boldsymbol{z}_B=T([f_1(B),\ldots,f_k(B)])$, where $T(\cdot)$ normalizes features such as code length, AST complexity, CFG depth, and current training behavior. A router $\rho_\omega(d\mid \boldsymbol{z}_B)$ then outputs a binary decision $d_B\in\{0,1\}$, where $d_B=0$ selects SFT and $d_B=1$ selects RFT. Appendix~\ref{sec:appendix_4} gives the architecture and optimization details.

The two update paths optimize different signals. For SFT-routed batches, the model minimizes the standard negative log-likelihood $\mathcal{L}_{\text{SFT}}(B)=-\sum_{(x_i,y_i)\in B}\log \pi_\theta(y_i\mid x_i)$. For RFT-routed batches, the model samples candidates, scores them with $r(x,\hat{y})$, and applies the RLOO objective described in Appendix~\ref{sec:RFT}. The routing design reflects a simple repair prior: many examples are best handled by learning common edit patterns directly, while harder examples benefit from symbolic feedback. Routing therefore preserves the efficiency and stability of SFT while using RFT selectively for examples where compiler-informed feedback is likely to help.

\subsection{Integrated Training Algorithm}

% Algorithm~\ref{alg:SynthFix} summarizes the training loop. At each batch, \TheName{} computes routing features, samples a route, and applies exactly one update path. This makes the comparison with SFT-only and RFT-only baselines controlled: all methods use the same base model, optimizer budget, data splits, and warmup schedule, while only the routing and symbolic feedback mechanisms differ.

Algorithm~\ref{alg:SynthFix} summarizes the training loop. For each batch, \TheName{} computes routing features, samples a route, and applies one update. This keeps comparisons with SFT-only and RFT-only baselines controlled: all methods share the same model, optimizer budget, splits, and warmup schedule, differing only in routing and symbolic feedback.

\begin{algorithm}[t]
\caption{Training Pipeline of \TheName{}}
\label{alg:SynthFix}
\begin{algorithmic}[1]
\REQUIRE Dataset $\mathcal{D}=\{(x_i,y_i)\}_{i=1}^{N}$, policy $\pi_\theta$, router $\rho_\omega$
\FOR{each batch $B \subset \mathcal{D}$}
  \STATE $\boldsymbol{z}_B \gets T([f_1(B),\ldots,f_k(B)])$ \COMMENT{Normalize routing features}
  \STATE $d_B \sim \rho_\omega(\cdot\mid\boldsymbol{z}_B)$ \COMMENT{0: SFT, 1: RFT}
  \IF{$d_B=0$}
    \STATE Compute supervised loss $\mathcal{L}_{\text{SFT}}(B)$
    \STATE Update $\theta$ with $\nabla \mathcal{L}_{\text{SFT}}(B)$
  \ELSE
    \STATE Sample candidates $\mathcal{C}_K(x_i)$ for each $x_i\in B$
    \STATE Score candidates with $r(x_i,\hat{y})=\boldsymbol{\lambda}^{\top}\boldsymbol{s}(x_i,\hat{y})$
    \STATE Compute reward-fine-tuning loss $\mathcal{L}_{\text{RFT}}(B)$
    \STATE Update $\theta$ with $\nabla \mathcal{L}_{\text{RFT}}(B)$
  \ENDIF
\ENDFOR
\end{algorithmic}
\end{algorithm}

\subsection{Symbolic-Guided Inference-Time Selection}

At inference time, \TheName{} samples $K$ candidate patches and ranks them using the same family of symbolic evidence used during training. Let $\hat{y}^{(0)}$ be the greedy decode and $\mathcal{C}_K(x)$ be the full candidate set. We first construct a valid subset $\mathcal{V}_K(x)=\{\hat{y}\in \mathcal{C}_K(x): \textsc{PassesAvailableValidityChecks}(x,\hat{y})\}$, where the checks may include parseability, compilation, public tests, or static security scans depending on the benchmark. Surviving candidates are scored by a selector $q(x,\hat{y})=\boldsymbol{\eta}^{\top}\boldsymbol{s}(x,\hat{y})$, and the best symbolic candidate is $\hat{y}^{\star}=\arg\max_{\hat{y}\in\mathcal{V}_K(x)}q(x,\hat{y})$. \TheName{} then applies a greedy floor: it returns $\hat{y}^{\star}$ only if it scores above the greedy decode, and otherwise returns $\hat{y}^{(0)}$. This selection step is leak-free because $q$ uses public tests and static signals only; held-out tests are reserved for final evaluation.

\section{Experimental Design}
\label{sec:experimental-design}

We evaluate \TheName{} on three execution- and security-grounded repair benchmarks across five modern code language models. This section summarizes the models, datasets, metrics, and research questions; full hyperparameters are in Appendix~\ref{sec:appendix_2}.

\subsection{Models and Baselines}

We fine-tune five code LLMs spanning 1.3B--7B parameters: DeepSeek-Coder-1.3B~\cite{guo2024deepseekcoder}, Llama-3.2-3B~\cite{dubey2024llama3}, Qwen3-4B~\cite{yang2025qwen3}, CodeLLaMA-7B~\cite{roziere2023code}, and StarCoder2-7B~\cite{lozhkov2024starcoder}. For each base model, we compare \TheName{} with SFT-only and RFT-only baselines. All methods are trained fresh from the same base model with LoRA and matched data, optimizer, and training budgets.

\subsection{Benchmarks and Evaluation Metrics}

We evaluate our framework on three execution- and security-grounded benchmarks: (1) \textbf{pyrepair}, a Python execution-repair benchmark we construct from MBPP~\cite{austin2021mbpp} (execution-validated bug injection for train/validation/test) augmented with QuixBugs~\cite{lin2017quixbugs} as a real-bug test set; (2) \textbf{CodeFlaws} for C ($\sim$4k defects, each shipping a test suite)~\cite{tan2017codeflaws}; and (3) \textbf{SVEN}~\cite{he2023sven}, a security-repair benchmark. We use a fixed random seed for reproducible train/validation/test splits across all experiments.

Repair quality is assessed primarily using \emph{deployable} metrics that measure whether a patch works: \textbf{functional pass@1} (the patched program compiles and passes its held-out test suite) on pyrepair and CodeFlaws, and \textbf{security-cleared rate} (a static analyzer, Semgrep, reports the vulnerability removed) on SVEN. Selection of best-of-$K$ candidates is leak-free: candidates are ranked using only public tests and static symbolic signals, scored on held-out tests, and constrained by a greedy floor so the selected patch can never regress below greedy decoding. We additionally report \textbf{CodeBLEU}~\cite{ren2020codebleu} as a diagnostic, noting that it saturates on these benchmarks and does not measure functional correctness.

All experiments were conducted on two NVIDIA A6000 GPUs.

\subsection{Research Questions}

Our evaluation is organized around four research questions:

\begin{description}
    \item[RQ1: End-to-End Repair Quality.] How does \TheName{} compare to SFT-only and RFT-only baselines on deployable repair metrics?
    \item[RQ2: Training- and Test-Time Modules.] How much do the adaptive router and symbolic best-of-$K$ selector contribute beyond the base policy?
    \item[RQ3: Generalization Across Defect Types.] Are gains consistent across the functional and security benchmark families we evaluate?
    \item[RQ4: Pipeline Ablation.] What is the contribution of the router, reward-driven training, and symbolic test-time selector to the final repair rate?
\end{description}

\section{Result Analysis}

This section reports end-to-end repair quality, then analyzes the router, the symbolic selector, benchmark-level generalization, pipeline ablations, and qualitative cases.

\subsection{Overall Performance (RQ1 \& RQ2)}

\noindent\textbf{End-to-end quality.} To answer \textbf{RQ1}, we evaluate \TheName{} on \emph{deployable} metrics: functional pass@1 on pyrepair and CodeFlaws, and security-cleared rate on SVEN. Table~\ref{tab:main_results} compares \TheName{} with SFT-only and budget-matched RFT-only baselines across five code LLMs. \TheName{} improves over SFT on every benchmark/model pair, with gains of $+12.9\%$ to $+26.6\%$ on pyrepair, $+20.5\%$ to $+53.6\%$ on CodeFlaws, and up to $+14.3\%$ in security clearance (ties occur at the ceiling). RFT alone stays close to SFT and sometimes regresses under greedy decoding, showing that training-time RL by itself is not sufficient for reliable deployable repair.

\begin{table*}[t]
    \centering
    \caption{Main deployable repair results across five base models. Functional pass@1 is reported for pyrepair and CodeFlaws, security-cleared rate for SVEN, and $\Delta$ is the relative gain of \TheName{} over SFT; best values are in \textbf{bold}.}
    \resizebox{0.95\linewidth}{!}{%
        \begin{tabular}{l r r r r | r r r r | r r r r}
            \toprule
            \multirow{2}{*}{Model} &
            \multicolumn{4}{c|}{pyrepair (Python, $n{=}115$)} &
            \multicolumn{4}{c|}{CodeFlaws (C, $n{=}389$)} &
            \multicolumn{4}{c}{SVEN (security, $n{=}16$)} \\
            \cmidrule(lr){2-5} \cmidrule(lr){6-9} \cmidrule(lr){10-13}
            & SFT & RFT & \TheName{} & $\Delta$ & SFT & RFT & \TheName{} & $\Delta$ & SFT & RFT & \TheName{} & $\Delta$ \\
            \midrule
            DeepSeek-1.3B   & 68.7 & 69.6 & \textbf{87.0} & +26.6\% & 12.3 & 10.5 & \textbf{15.2} & +22.9\% & 87.5  & 87.5 & \textbf{100}  & +14.3\% \\
            Llama-3.2-3B    & 73.9 & 73.9 & \textbf{85.2} & +15.3\% & 14.4 & 13.6 & \textbf{22.1} & +53.6\% & 100   & 87.5 & 100           & --      \\
            Qwen3-4B-Base   & 80.9 & 80.9 & \textbf{91.3} & +12.9\% & 18.8 & 15.9 & \textbf{22.6} & +20.5\% & 87.5  & 87.5 & \textbf{93.8} & +7.1\%  \\
            CodeLLaMA-7B    & 72.2 & 75.7 & \textbf{81.7} & +13.3\% & 12.9 & 12.6 & \textbf{18.8} & +46.0\% & 93.8  & 93.8 & 93.8          & --      \\
            StarCoder2-7B   & 79.1 & 76.5 & \textbf{93.0} & +17.6\% & 15.7 & 15.7 & \textbf{22.6} & +44.3\% & 87.5  & 81.2 & \textbf{100}  & +12.5\% \\
            \bottomrule
        \end{tabular}%
    }
    \label{tab:main_results}
\end{table*}

\noindent\textbf{Module contributions.} Having established the end-to-end gain, we next answer \textbf{RQ2} by separating the \emph{training-time} and \emph{test-time} modules.

\begin{table}[t]
\centering
\caption{Controlled module diagnostics for DeepSeek-1.3B ($K{=}16$). The router comparison isolates adaptive training on CodeFlaws, while the selector comparison measures greedy, random, symbolic best-of-$K$, and oracle outcomes; values are solved@1 and $\Delta$ is relative to the same benchmark's greedy floor.}
\label{tab:module_effectiveness}
\resizebox{\columnwidth}{!}{%
\begin{tabular}{l l l r r}
\toprule
Module & Benchmark & Variant & Solved@1 & $\Delta$ \\
\midrule
\multirow{2}{*}{Router}
& CodeFlaws & Fixed schedule (no router) & 46 (11.8\%) & --- \\
& CodeFlaws & \TheName{} router & \textbf{50 (12.9\%)} & +4 \\
\midrule
\multirow{7}{*}{Selector}
& CodeFlaws & Greedy floor & 50 (12.9\%) & --- \\
& CodeFlaws & Random pick from $K$ & 41 (10.5\%) & $-9$ \\
& CodeFlaws & Symbolic best-of-$K$ & \textbf{66 (17.0\%)} & +16 \\
& CodeFlaws & Oracle@$K$ (upper bound) & 86 (22.1\%) & +36 \\
& pyrepair & Greedy floor & 79 (68.7\%) & --- \\
& pyrepair & Symbolic best-of-$K$ & \textbf{100 (87.0\%)} & +21 \\
& pyrepair & Oracle@$K$ (upper bound) & 104 (90.4\%) & +25 \\
\bottomrule
\end{tabular}%
}
\end{table}

Table~\ref{tab:module_effectiveness} separates the two modules in controlled diagnostics. On CodeFlaws, the adaptive router improves the trained policy over a fixed no-router schedule (50 vs.\ 46 solved). The larger gain comes at test time: symbolic best-of-$K$ raises the same policy from 50 to 66 solved on CodeFlaws, and from 79 to 100 solved on pyrepair. Randomly choosing from the CodeFlaws candidate pool is worse than greedy, so the improvement is not merely from sampling more patches; it comes from ranking candidates with symbolic signals and a greedy floor.

\begin{figure}[t]
    \centering
    \includegraphics[width=\columnwidth]{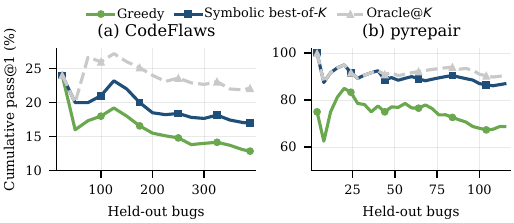}
    \caption{Cumulative test-time selection for DeepSeek-1.3B ($K{=}16$). Symbolic best-of-$K$ stays above greedy on CodeFlaws and pyrepair, while oracle@$K$ shows the remaining reachable headroom.}
    \label{fig:module_analysis}
\end{figure}

Figure~\ref{fig:module_analysis} visualizes this test-time module on the same trained policy: symbolic best-of-$K$ consistently outperforms greedy decoding and approaches the oracle@$K$ ceiling as more held-out bugs are evaluated. Panel~(a) shows the pattern on CodeFlaws (+16 solved at $n{=}389$); panel~(b) shows the same trend on pyrepair.

% \noindent\textbf{Selection efficiency.} Because Table~\ref{tab:module_effectiveness} isolates the selector on a single model, Table~\ref{tab:selection_efficiency} generalizes the analysis to all five base models by adding the oracle@$K$ ceiling and reporting how much of that reachable headroom the symbolic selector captures. Selection recovers $81$--$94\%$ of the oracle gap on pyrepair (avg.\ $87.4\%$), $38$--$70\%$ on the harder CodeFlaws C defects (avg.\ $53.5\%$), and $83.3\%$ of aggregate security headroom on SVEN where nonzero headroom remains. This confirms that the test-time benefit is a consistent, model-agnostic property rather than an artifact of one base model, while the residual gap to oracle quantifies the headroom that stronger candidate generation could still unlock.

\noindent\textbf{Selection efficiency.} Because Table~\ref{tab:module_effectiveness} isolates the selector on a single model, Table~\ref{tab:selection_efficiency} generalizes to five base models by adding the oracle@$K$ ceiling and reporting how much headroom the symbolic selector captures. Selection recovers $81$--$94\%$ of the oracle gap on pyrepair (avg.\ $87.4\%$), $38$--$70\%$ on harder CodeFlaws C defects (avg.\ $53.5\%$), and $83.3\%$ of aggregate security headroom on SVEN where nonzero headroom remains. This confirms the test-time benefit is consistent and model-agnostic, not an artifact of one base model, while the residual oracle gap quantifies headroom stronger candidate generation could still unlock.

The captured-headroom view is important because it separates two failure modes. If oracle@$K$ is low, the generator is not producing enough correct candidates; if oracle@$K$ is high but best-of-$K$ is low, the selector is failing to identify them. \TheName{} mainly narrows the second gap, especially on pyrepair and SVEN, while CodeFlaws shows that harder C defects still need both better candidates and sharper ranking.

\begin{table*}[t]
    \centering
    \caption{Test-time selection efficiency across all five base models ($K{=}16$). For each benchmark, we compare the greedy floor, symbolic best-of-$K$, and oracle@$K$ ceiling; Captured is the fraction of oracle headroom recovered by symbolic selection, and -- indicates no oracle headroom.}
    \resizebox{\linewidth}{!}{%
        \begin{tabular}{l r r r r | r r r r | r r r r}
            \toprule
            \multirow{2}{*}{Model} &
            \multicolumn{4}{c|}{pyrepair (Python, $n{=}115$)} &
            \multicolumn{4}{c|}{CodeFlaws (C, $n{=}389$)} &
            \multicolumn{4}{c}{SVEN (security, $n{=}16$)} \\
            \cmidrule(lr){2-5} \cmidrule(lr){6-9} \cmidrule(lr){10-13}
            & Greedy & Best-of-$K$ & Oracle@$K$ & Captured & Greedy & Best-of-$K$ & Oracle@$K$ & Captured & Greedy & Best-of-$K$ & Oracle@$K$ & Captured \\
            \midrule
            DeepSeek-1.3B   & 68.7 & 87.0 & 90.4 & 84.0\% & 12.3 & 15.2 & 19.3 & 40.7\% & 87.5 & 100 & 100 & 100\% \\
            Llama-3.2-3B    & 73.9 & 85.2 & 87.8 & 81.3\% & 14.4 & 22.1 & 27.2 & 60.0\% & 100 & 100 & 100 & -- \\
            Qwen3-4B-Base   & 80.9 & 91.3 & 93.0 & 85.7\% & 18.8 & 22.6 & 28.8 & 38.5\% & 87.5 & 93.8 & 100 & 50.0\% \\
            CodeLLaMA-7B    & 72.2 & 81.7 & 82.6 & 91.7\% & 12.9 & 18.8 & 21.3 & 69.7\% & 93.8 & 93.8 & 93.8 & -- \\
            StarCoder2-7B   & 79.1 & 93.0 & 93.9 & 94.1\% & 15.7 & 22.6 & 27.5 & 58.7\% & 87.5 & 100 & 100 & 100\% \\
            \midrule
            Average         & 75.0 & 87.6 & 89.5 & 87.4\% & 14.8 & 20.3 & 24.8 & 53.5\% & 91.3 & 97.5 & 98.8 & 83.3\% \\
            \bottomrule
        \end{tabular}%
    }
    \label{tab:selection_efficiency}
\end{table*}

\subsection{RQ3: Generalization Across Defect Types}

We next examine whether the gains persist across functional Python repair, functional C repair, and security repair.

\begin{table}[t]
\centering
\caption{Category-level summary averaged over the five base models. The table groups the benchmark suite by repair property and reports SFT, RFT, \TheName{}, oracle@$K$ sampled-candidate ceilings, and the fraction of oracle headroom captured.}
\label{tab:category_analysis}
\resizebox{\columnwidth}{!}{%
\begin{tabular}{l r r r r r r}
\toprule
Benchmark category & $n$ & SFT & RFT & \TheName{} & Oracle@$K$ & Captured \\
\midrule
pyrepair (Python) & 115 & 75.0 & 75.3 & \textbf{87.6} & 89.5 & 87.4\% \\
CodeFlaws (C) & 389 & 14.8 & 13.7 & \textbf{20.3} & 24.8 & 53.5\% \\
SVEN (security) & 16 & 91.3 & 87.5 & \textbf{97.5} & 98.8 & 83.3\% \\
\midrule
\multicolumn{2}{l}{Avg.\ absolute gain over SFT} & -- & $-1.5$ & \textbf{+8.1} & -- & -- \\
\bottomrule
\end{tabular}%
}
\end{table}

Table~\ref{tab:category_analysis} gives a category-level view and adds the oracle@$K$ ceiling. \TheName{} improves the average score in every category. The captured-headroom column shows that pyrepair and SVEN are close to their sampled-candidate ceilings, while CodeFlaws leaves more unresolved oracle headroom. Because the current SVEN subset is dominated by CWE-89 (SQL injection), we report the benchmark-level security result rather than a fine-grained multi-CWE breakdown.

\begin{figure}[b]
    \centering
    \includegraphics[width=\columnwidth]{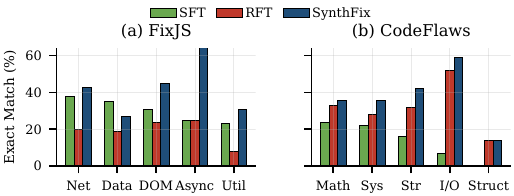}
    \caption{Auxiliary exact-match analysis by functionality group in the FixJS/CodeFlaws diagnostic setting. This figure is retained for qualitative context; main claims use deployable metrics.}
    \label{fig:functionality_analysis}
\end{figure}

\noindent\textbf{Auxiliary functionality diagnostic.} Figure~\ref{fig:functionality_analysis} reports a functionality-group analysis from the earlier FixJS/CodeFlaws exact-match setting. This figure is included only to give qualitative context about the kinds of programming patterns affected by repair models; it is not used for the main deployable-metric claims, which are supported by Table~\ref{tab:main_results}, Table~\ref{tab:category_analysis}, and Figure~\ref{fig:module_analysis}.

\subsection{Pipeline Ablation (RQ4)}

Finally, we ablate the main components of the \TheName{} pipeline. Because these controlled variants are run on CodeFlaws with DeepSeek-1.3B for tractability, Table~\ref{tab:rq5_ablation_single} should be read as a focused component analysis rather than a full five-model sweep.

\begin{table}[t]
\centering
\caption{Component ablation on CodeFlaws functional pass@1 (DeepSeek-1.3B, $n{=}389$, $K{=}16$). Rows remove or replace one major pipeline component while keeping the split and evaluation protocol fixed; best non-oracle result is in \textbf{bold}.}
\label{tab:rq5_ablation_single}
\resizebox{\columnwidth}{!}{%
\begin{tabular}{l l r}
\toprule
Variant & Selection rule & Solved@1 \\
\midrule
SFT-only baseline & Greedy & 45 (11.6\%) \\
RFT-only baseline & Greedy & 41 (10.5\%) \\
\midrule
No router (fixed schedule) & Greedy & 46 (11.8\%) \\
No router (fixed schedule) & Symbolic best-of-$K$ & 64 (16.5\%) \\
\midrule
\TheName{} without test-time selector & Greedy & 50 (12.9\%) \\
\TheName{} with unranked candidates & Random pick from $K$ & 41 (10.5\%) \\
\TheName{} full pipeline & Symbolic best-of-$K$ & \textbf{66 (17.0\%)} \\
\midrule
Oracle@$K$ upper bound & Held-out oracle & 86 (22.1\%) \\
\bottomrule
\end{tabular}%
}
\end{table}

Table~\ref{tab:rq5_ablation_single} makes three points. First, RFT-only training under greedy decoding does not improve over SFT (41 vs.\ 45 solved), so symbolic rewards alone are not enough. Second, replacing the router with a fixed schedule lowers the greedy policy (46 vs.\ 50 solved), while symbolic best-of-$K$ still recovers many repairs (64 solved). Third, sampling without symbolic ranking is insufficient: random pick falls to 41 solved, whereas the full selector solves 66. The oracle@$K$ value of 86 shows that stronger generation and ranking still leave headroom.

This ablation also clarifies the role of the greedy floor. The selector is not allowed to trade a known greedy repair for an unsupported sampled candidate unless the symbolic evidence justifies it. That design is conservative, but it is appropriate for repair: deployability depends more on avoiding regressions than on maximizing textual diversity.

\subsection{Qualitative Case Study}

\begin{figure*}[t]
    \centering
    \includegraphics[width=\textwidth]{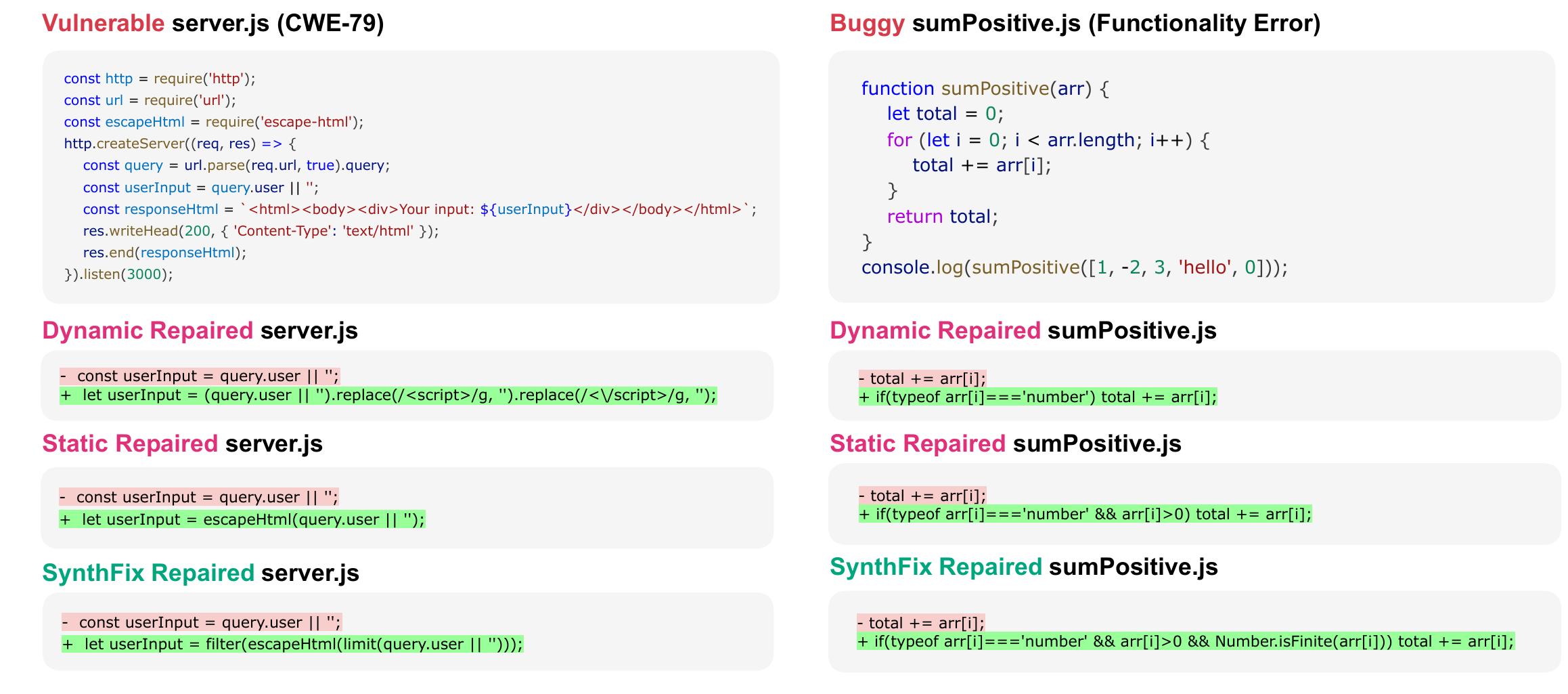}
    \caption{Representative repairs from SFT, RFT, and \TheName{}. The examples show a security sanitization case and a logic repair where symbolic evidence helps select the deployable patch.}
    \label{fig:case-diff}
\end{figure*}

We finish with representative repair cases. Figure~\ref{fig:case-diff} compares patches from SFT, RFT, and \TheName{} for two scenarios and illustrates why symbolic feedback is useful even when surface-form metrics are ambiguous.

\paragraph{Case Study 1: Cross-Site Scripting (CWE-79).}
As shown in the left panel of Figure~\ref{fig:case-diff}, a JavaScript web server is vulnerable to script injection. The \textbf{SFT} model applies a naive fix that only strips \texttt{<script>} tags, while the \textbf{RFT} model applies \texttt{escapeHtml(...)} but performs no other checks. By combining the pattern learning of SFT with symbolic validation, \TheName{} generates a multi-layered defense:
\begin{lstlisting}[language=JavaScript, basicstyle=\small\ttfamily, breaklines=true]
let userInput = filter(escapeHtml(limit(query.user || '')));
\end{lstlisting}
This repair composes length limiting, HTML escaping, and allow-list filtering, which is more robust than either standalone baseline.

\paragraph{Case Study 2: Array Summation Logic Error.}
The right panel of Figure~\ref{fig:case-diff} shows a function that incorrectly sums array elements. The \textbf{SFT} model adds a partial type check but still allows negative numbers. The \textbf{RFT} model produces a correct but verbose condition with a redundant check. \TheName{} synthesizes the concise repair:
\begin{lstlisting}[language=JavaScript, basicstyle=\small\ttfamily, breaklines=true]
if (typeof arr[i] === 'number' && arr[i] > 0) total += arr[i];
\end{lstlisting}
This succinct condition enforces the intended semantics without unnecessary checks, illustrating how the adaptive pipeline can produce repairs that are both compact and correct.

\section{Discussion}

Our results show that supervised learning, symbolic reinforcement feedback, and symbolic test-time selection play complementary roles in automated program repair. SFT provides a stable foundation for common repair patterns, while RFT targets examples that need stronger semantic evidence. This mirrors a practical repair workflow: known fixes can often be applied directly, but harder bugs require additional feedback from tests, analyzers, or compiler behavior.

\noindent\textbf{Adaptive training.} The module analysis suggests that repair benefits from allocating training effort based on code-level difficulty. Some bugs are formulaic and can be learned well through imitation, while others require feedback-driven exploration. In the controlled CodeFlaws study, the router improves the trained policy over a fixed schedule, showing that a single uniform objective is not always the best use of the training budget.

\noindent\textbf{Inference-time validation.} \TheName{} also highlights the importance of symbolic feedback after training. The largest deployable gains come from combining adaptive symbolic training with symbolic best-of-$K$ selection (Table~\ref{tab:main_results}, Figure~\ref{fig:module_analysis}). The component ablation in Table~\ref{tab:rq5_ablation_single} shows that neither reward-driven training nor candidate sampling is sufficient alone; the strongest result comes from coupling generation with evidence-based selection.

\noindent\textbf{Deployment implication.} This matters because stronger code LLMs may produce more plausible candidate patches, but plausibility is not a deployment criterion. A repair workflow still needs evidence that a patch compiles, preserves intended behavior, and removes the vulnerability. \TheName{} treats this evidence as part of the decision process rather than only an after-the-fact evaluation metric, making the repair pipeline more aligned with deployable software maintenance.

\section{Related Work}

Our work combines automated program repair, neural code generation, reinforcement learning, and compiler-informed analysis. We organize prior work around the same design choices as \TheName{}: SFT-based repair, feedback-driven refinement, symbolic analysis, and inference-time validation.

\subsection{Neural Program Repair with SFT}
Neural and LLM-based methods have become central to automated code repair~\cite{allamanis2018survey, chen2021evaluating, li2024machines, bansal2023modeling, zhang2022leveraging, zhang2025leveraging, zhang2025enhancing, richter2022neural, feng2020codebert}. Most use Supervised Fine-Tuning (SFT) to learn mappings from buggy to fixed code from large repair datasets~\cite{yin2017syntactic, hayati2018retrieval, habib2019neural, li2019improving, gupta2019neural, allamanis2021self, jiang2023impact}. While effective for common syntactic patterns, these methods can still generate plausible patches that are functionally incorrect or fail to remove the vulnerability~\cite{berabi2021tfix, huang2023empirical, pailoor2024semantic, zhang2022pre, kou2024large, ziems2021security}.

\subsection{Reinforcement Learning for Code}
% Reward Fine-Tuning (RFT) and reinforcement learning address this limitation by optimizing models with dynamic feedback, such as unit tests or security checks~\cite{le2022coderl, shojaee2023execution, fang2025dpo}. PPO is commonly used for such iterative refinement~\cite{schulman2017proximal}, and recent work also incorporates multi-step symbolic verification signals~\cite{zhang2025codegrad}. However, these methods can be computationally expensive and slow to converge, limiting their practicality for large-scale repair~\cite{huang2023empirical, shi2023towards}.

Reward Fine-Tuning (RFT) and reinforcement learning address this limitation by optimizing models with dynamic feedback, such as unit tests or security checks~\cite{le2022coderl, shojaee2023execution, fang2025dpo}. PPO is commonly used for iterative refinement~\cite{schulman2017proximal}, and recent work incorporates multi-step symbolic verification signals~\cite{zhang2025codegrad}. However, these methods can be computationally expensive and slow to converge, limiting practicality for large-scale repair~\cite{huang2023empirical, shi2023towards}.

\subsection{Symbolic and Compiler-Informed Analysis}
Compiler-informed repair methods use representations such as ASTs and CFGs to capture structural and semantic properties of code~\cite{shi2020pathpair2vec, wu2022turn, jiang2018shaping, klieber2021automated, mandal2018generic, wang2021syncobert}. These signals are precise and useful, but many traditional systems depend on predefined heuristics or manually crafted templates, which can limit flexibility on diverse real-world vulnerabilities~\cite{zhang2022astro}.

\subsection{Inference-Time Validation and Selection}
Recent repair pipelines increasingly separate candidate generation from candidate acceptance when execution or validation feedback is available~\cite{gupta2020synthesize,le2022coderl,shojaee2023execution,kulsum2024case}. This distinction is important because the highest-probability patch may not compile, pass tests, or remove the vulnerability. Generate-and-rank workflows therefore use symbolic evidence to filter invalid patches, score candidates with public tests or static analyzers, and compare alternatives against a greedy baseline. Prior work on developer validation of LLM-generated code similarly shows that explicit checking is needed beyond surface plausibility~\cite{tang2024developer}. Larger candidate pools can increase oracle headroom, but can also introduce regressions if the final choice is random or based only on likelihood, motivating the greedy floor and symbolic reranker in \TheName{}.

% \TheName{} differs from methods that use symbolic checks only as a final filter. Symbolic evidence appears in three places: it shapes reward for feedback-driven updates, supplies routing features for deciding when such updates are useful, and ranks candidate patches at inference time. The goal is not to replace neural repair with a hand-written analyzer, but to keep neural generation anchored to evidence that corresponds to deployable repair behavior.

\noindent\textbf{Neuro-symbolic positioning.} \TheName{} differs from methods that use symbolic checks only as a final filter. Symbolic evidence appears in three places: it shapes reward for feedback-driven updates, supplies routing features for deciding when such updates are useful, and ranks candidate patches at inference time. The goal is not to replace neural repair with a hand-written analyzer, but to keep neural generation anchored to evidence that corresponds to deployable repair behavior.
\section{Conclusion and Future Work}

% We introduced \TheName{}, an adaptive neuro-symbolic framework for automated vulnerability repair. By routing training between SFT and RFT and reusing symbolic evidence for best-of-$K$ selection, \TheName{} aligns neural patch generation with compiler-, analyzer-, and test-facing repair signals. Across five code LLMs and three benchmarks, it improves over SFT-only and RFT-only baselines; module analyses show that the gains come from the full training-and-selection pipeline rather than sampling alone.

We introduced \TheName{}, an adaptive neuro-symbolic framework for automated vulnerability repair. By routing training between SFT and RFT and reusing symbolic evidence for best-of-$K$ selection, \TheName{} aligns neural patch generation with compiler-, analyzer-, and test-facing repair signals. Across five code LLMs and three benchmarks, it improves over SFT-only and RFT-only baselines; module analyses show that the gains come from the full training-and-selection pipeline rather than sampling alone. These results suggest that effective repair systems should not rely only on larger generators, but should also use symbolic evidence to decide when models should learn from feedback, how candidate patches should be ranked, and which repair is safe enough to deploy. More broadly, \TheName{} shows that neural generation and symbolic validation can be integrated throughout the repair pipeline rather than treated as separate stages.

\section*{Limitations}

A primary limitation is the Router Model's reliance on static code features (e.g., AST complexity, CFG depth). These features are inexpensive and useful, but they miss dynamic runtime behavior that could better guide complex repairs involving subtle execution logic. In addition, our evaluation uses established academic benchmarks. While standard for this area, they may not reflect the scale and dependencies of large industrial systems. Thus, \TheName{}'s performance in such settings remains future work.

\paragraph{Threats to validity.} \emph{Construct:} surface-overlap metrics (CodeBLEU/exact match) saturate and do not track whether a patch runs (Table~\ref{tab:rq1_comparison}), so our claims rely on execution- and security-grounded metrics (functional pass@1, Semgrep-verified clearance). \emph{Internal:} we train SFT, RFT, and \TheName{} fresh from the same base model with identical splits, optimizer, and budget, and keep best-of-$K$ selection leak-free through a public-test/static-signal-only reranker with a greedy floor, so gains cannot come from held-out test access; the module analysis separates the router and symbolic selector from larger candidate pools. \emph{External:} our deepest ablations use DeepSeek-1.3B on CodeFlaws for tractability, but the headline comparison spans five model families (1.3B--7B) and three benchmarks, with \TheName{} either improving over SFT or tying at a ceiling in every benchmark/model cell (Table~\ref{tab:main_results}); generalization to multi-file industrial repositories remains future work.

\section*{Ethical Considerations}

While \TheName{} aims to improve software security, we note the risks common to any automated repair tool. Chief among them is a false sense of security: a generated patch may be incomplete or introduce subtle runtime errors, so over-reliance could reduce manual auditing. We therefore intend \TheName{} as an assistive tool that augments---rather than replaces---developer oversight, and advocate deploying it only within workflows that mandate testing and human review of all generated patches.

\paragraph{Use of AI Assistants.} The authors used AI assistants (LLMs) for minor editorial assistance, including language polishing and LaTeX formatting. All research ideation, experimental design, implementation, analysis, and claims are the authors' own.

\bibliography{custom}

\appendix

\section{Technical Background for \TheName{}}
\label{sec:appendix_1}

This appendix provides implementation details that support the main paper. We briefly define the SFT/RFT objectives used by \TheName{}, then document the model configuration, reward signals, inference-time selector, benchmark construction, and auxiliary diagnostics.

\subsection{Supervised Fine-Tuning (SFT)}
\label{sec:SFT}

Supervised Fine-Tuning is a standard technique for adapting pre-trained models to specific tasks. Given a dataset of buggy code snippets $x_i$ and their corresponding ground-truth fixes $y_i$, SFT optimizes the model's parameters $\theta$ by minimizing the negative log-likelihood of the target sequences: $\mathcal{L}_{\text{SFT}}(\theta) = - \sum_{i} \log P_{\theta}(y_i \mid x_i)$.

In \TheName{}, SFT provides the stable imitation pathway for learning common repair patterns. It also defines the warmup checkpoint used by all methods before the reward-driven stage, ensuring that SFT-only, RFT-only, and \TheName{} share the same supervised starting point.

\subsection{Reward Fine-Tuning (RFT)}
\label{sec:RFT}

Reward Fine-Tuning uses reinforcement learning to optimize a model based on feedback from a reward function, which is particularly useful when a ground-truth output is not available or when optimizing for complex criteria beyond surface-level similarity. We instantiate RFT with a variance-reduced REINFORCE objective using a \emph{leave-one-out} (RLOO) baseline. For each prompt $x$, the policy samples a group of $k$ candidate patches $\{\hat{y}^{(1)},\dots,\hat{y}^{(k)}\}$; each candidate is scored by the symbolic reward $r(\hat{y}^{(j)})$, and its advantage is computed against the mean reward of the \emph{other} candidates in the group, $\hat{A}^{(j)} = r(\hat{y}^{(j)}) - \tfrac{1}{k-1}\sum_{l\neq j} r(\hat{y}^{(l)})$. To keep the policy from drifting away from the fluent SFT reference and collapsing, we add a KL-divergence penalty to the frozen SFT reference policy $\pi_{\text{ref}}$, giving the compact objective $\mathcal{L}_{\text{RFT}}(\theta)=-\mathbb{E}[\hat{A}^{(j)}\log \pi_\theta(\hat{y}^{(j)}\mid x)]+\beta_{\text{KL}}\mathrm{KL}(\pi_\theta\|\pi_{\text{ref}})$.
This avoids training a separate value network while retaining low-variance updates. In \TheName{}, RFT is the feedback-driven pathway: it exposes candidate patches to symbolic feedback about syntax, compilation, public tests, security scans, and repair-oriented edits. Exact RL hyperparameters are given in Appendix~\ref{sec:appendix_2}.

\subsection{Symbolic Representations for Reward and Routing}
\label{sec:ast-cfg}

Our framework integrates symbolic reasoning by leveraging compiler-informed code representations, primarily Abstract Syntax Trees (ASTs), Control Flow Graphs (CFGs), and data-flow features.

An \textbf{Abstract Syntax Tree (AST)} represents hierarchical syntactic structure. A \textbf{Control Flow Graph (CFG)} models possible execution paths, and data-flow features summarize how values are produced and consumed.

These representations serve a dual purpose in \TheName{}:
\begin{enumerate}
    \item \textbf{For adaptive routing:} Compiler-derived metrics such as code length, AST complexity, and CFG depth form the feature vector $f_B$ used by the Router Model to choose between SFT and RFT updates.
    \item \textbf{For reward and selection:} The same families of evidence are used to score candidate patches during RFT and to rank sampled candidates at inference time. We assess a candidate patch $\hat{y}$ with a composite reward over static structure, lint/compile validity, security scanning, repair-effect signals, and, when available, public execution tests:
    $r(\hat{y}) = \sum_c \lambda_c r_c(\hat{y})$.
    Each component provides distinct feedback, and the exact instantiation used in the experiments is described in Appendix~\ref{sec:appendix_reward}.
\end{enumerate}

\section{Detailed Experimental Setup}
\label{sec:appendix_2}

This appendix provides the model configurations, two-stage training recipe, symbolic reward implementation, and inference-time selection procedure used to evaluate \TheName{}, supplementing Section~\ref{sec:experimental-design}. All settings are identical across the five base models unless noted in Table~\ref{tab:appendix_modelcfg}; SFT-only, RFT-only, and \TheName{} use the same splits, LoRA setup, optimizer budget, and decoding budget so that comparisons are controlled.

\subsection{Repair Agent and Parameter-Efficient Fine-Tuning}

Each Repair Agent is the corresponding pre-trained base model (DeepSeek-Coder-1.3B, Llama-3.2-3B, Qwen3-4B-Base, CodeLLaMA-7B, StarCoder2-7B), adapted with \textbf{Low-Rank Adaptation (LoRA)} rather than full fine-tuning. We use a LoRA rank of $r{=}16$ with scaling $\alpha{=}32$ and dropout $0.05$, applied to the attention projections (\texttt{q\_proj}, \texttt{k\_proj}, \texttt{v\_proj}, \texttt{o\_proj}); base weights stay frozen. This keeps the trainable footprint below $1\%$ of model parameters and lets all five models---including the 7B variants---fit on two NVIDIA A6000 (48\,GB) GPUs. Inputs are tokenized with the model's native tokenizer using left-side padding and a maximum sequence length of $512$ tokens. Per-model batch sizes and decode lengths are listed in Table~\ref{tab:appendix_modelcfg}.

\begin{table}[t]
    \centering
    \small
    \caption{Per-model configuration and training budget. All models use the same LoRA rank, optimizer, four-epoch schedule, and $K{=}16$ decoding; gradient checkpointing (GC) is enabled for larger models.}
    \setlength{\tabcolsep}{5pt}
    \begin{tabular}{l c c c}
        \toprule
        Base model & Params & Batch size & GC \\
        \midrule
        DeepSeek-Coder-1.3B & 1.3B & 16 & --- \\
        Llama-3.2-3B        & 3B   & 8  & --- \\
        Qwen3-4B-Base       & 4B   & 8  & \checkmark \\
        CodeLLaMA-7B        & 7B   & 4  & \checkmark \\
        StarCoder2-7B       & 7B   & 4  & \checkmark \\
        \bottomrule
    \end{tabular}
    \label{tab:appendix_modelcfg}
\end{table}

\subsection{Two-Stage Training Recipe}

\TheName{} is trained \emph{fresh} from each base model in two stages, for a total of four epochs:

\begin{description}
    \item[Stage 1 --- SFT warmup (2 epochs).] We first minimize the standard supervised loss $\mathcal{L}_{\text{SFT}}$ (Appendix~\ref{sec:SFT}) on the (buggy, fixed) pairs to establish a fluent repair prior. The same warmed-up checkpoint defines the SFT-only baseline and the frozen reference policy $\pi_{\text{ref}}$ used by the KL anchor in Stage 2.
    \item[Stage 2 --- router-gated symbolic RL (2 epochs).] We then continue training with the router-gated RLOO objective of Appendix~\ref{sec:RFT}. For each prompt the policy samples a group of $k{=}2$ candidates; advantages use the leave-one-out baseline and the update is regularized toward $\pi_{\text{ref}}$.
\end{description}

The Repair Agent's adapters are optimized with \textbf{AdamW} (learning rate $2\times10^{-4}$, weight decay $0.01$, gradient-norm clipping at $1.0$). During Stage 2, RL rollouts are decoded with temperature $0.95$, top-$p$ $0.95$, and a no-repeat $3$-gram constraint to encourage candidate diversity; the RL term and the KL penalty use weights $\beta_{\text{RL}}{=}0.12$ and $\beta_{\text{KL}}{=}0.12$. The Router Model (Appendix~\ref{sec:appendix_4}) is optimized concurrently with a separate Adam optimizer at learning rate $1\times10^{-3}$. The RFT-only baseline uses the same Stage-2 objective but routes every batch through the RL pathway (no adaptive router). Final evaluation is always performed with the held-out test split and the leak-free selector described below.

\subsection{Symbolic Reward Implementation}
\label{sec:appendix_reward}

The composite reward $r(\hat{y})=\sum_c \lambda_c\, r_c(\hat{y})$ aggregates complementary symbolic signals, each normalized to $[0,1]$:

\begin{description}
    \item[$r_{\text{AST}}$ (syntactic soundness).] A parse of the candidate; full credit for a well-formed AST and a steep penalty for non-parseable code.
    \item[$r_{\text{CFG}}$ / $r_{\text{DFG}}$ (structural/data-flow fidelity).] Graph-similarity scores between the candidate and reference control-flow and data-flow graphs, combining node, edge, and path overlap.
    \item[$r_{\text{lint}}$ (static validity).] Linter/compiler-frontend signal rewarding code that passes static checks without warnings.
    \item[$r_{\text{Semgrep}}$ (security).] A Semgrep scan with a custom ruleset; the score starts at $100$ and is decremented for each matched vulnerability pattern.
    \item[$r_{\text{exec}}$ (execution).] When an executable suite is available, the fraction of \emph{public/visible} tests the patched program passes (held-out tests are never used in the reward).
    \item[$r_{\text{repair-effect}}$.] A reference-free term that rewards repair-directed edits (added guards, bounds checks, sanitizers, error handling) relative to the buggy input.
\end{description}

On execution benchmarks, the reward gives the largest weight to public-test execution and repair-effect signals, while retaining parse, data-flow, lint, CFG, AST, and surface-similarity terms. On static/security settings, execution is unavailable, so the reward relies on parse/lint/structure, Semgrep, and similarity signals. Reward computation is fully procedural (no learned reward model), which makes it deterministic and inexpensive to reproduce; the focused component ablation in the main text tests the pipeline modules rather than trying to rank individual reward weights.

\subsection{Inference-Time Best-of-$K$ Selection}
\label{sec:appendix_infer}

At decoding time \TheName{} samples $K{=}16$ candidates per bug using one greedy decode plus sampled decodes spanning temperatures $0.2$--$1.0$ with top-$p$ $0.95$. Candidates that fail available hard validity checks---such as parsing, compilation, or public/visible tests---are discarded before ranking. Surviving candidates are scored by a lightweight reranker over symbolic features: parseability, lint/compile status, CFG/DFG similarity, Semgrep score, public-test pass rate when available, candidate temperature, and average token log-probability. The top-ranked candidate is returned subject to a \textbf{greedy floor}: if no sampled candidate outranks the greedy decode, the greedy decode is kept. Crucially, ranking consults only public tests and static signals; held-out tests are reserved solely for final scoring. Algorithm~\ref{alg:infer} summarizes the procedure.

\begin{algorithm}[t]
\caption{Symbolic-Guided Inference-Time Selection}
\label{alg:infer}
\begin{algorithmic}[1]
\REQUIRE bug $x$, policy $\pi_\theta$, reranker $g$, budget $K$
\STATE $\hat{y}_0 \gets \textsc{GreedyDecode}(\pi_\theta, x)$
\STATE $\mathcal{C} \gets \{\hat{y}_0\} \cup \textsc{SampleK}(\pi_\theta, x, K{-}1)$ \COMMENT{mixed temps}
\STATE $\mathcal{V} \gets \{\hat{y}\in\mathcal{C}: \textsc{PassesAvailableValidityChecks}(\hat{y})\}$
\IF{$\mathcal{V} = \emptyset$}
  \STATE \textbf{return} $\hat{y}_0$ \COMMENT{greedy floor}
\ENDIF
\STATE $\hat{y}^{\star} \gets \arg\max_{\hat{y}\in\mathcal{V}} g(\phi(\hat{y}))$ \COMMENT{rank by symbolic features}
\STATE \textbf{return} $g(\phi(\hat{y}^{\star})) \geq g(\phi(\hat{y}_0))\ ?\ \hat{y}^{\star} : \hat{y}_0$
\end{algorithmic}
\end{algorithm}

\section{Benchmark Construction and Statistics}
\label{sec:appendix_bench}

This appendix details how each benchmark is built and split. All splits use a fixed random seed for reproducibility, and we deduplicate across splits to prevent train/test leakage.

\begin{description}
    \item[pyrepair (Python, execution).] We construct pyrepair from two sources. For training/validation/test we apply \emph{execution-validated bug injection} to MBPP~\cite{austin2021mbpp} problems: we introduce small, behavior-changing mutations (operator swaps, off-by-one edits, boundary and conditional perturbations) and keep only those whose buggy version fails at least one of the problem's tests while the original passes, yielding (buggy, fixed) pairs each shipping an executable test suite. To measure transfer to \emph{naturally occurring} defects, the held-out test set additionally includes QuixBugs~\cite{lin2017quixbugs} programs ($n{=}40$ classic algorithmic bugs); functional pass@1 is reported on the combined held-out set ($n{=}115$).
    \item[CodeFlaws (C, execution).] We use CodeFlaws~\cite{tan2017codeflaws}, $\sim$4k real defects mined from Codeforces submissions, each accompanied by a test suite. After filtering items that fail to build in our sandbox, the held-out test set contains $n{=}389$ defects. Functional pass@1 requires the patched program to compile and pass its held-out tests.
    \item[SVEN (security).] We use the SVEN~\cite{he2023sven} security-repair corpus of real vulnerable/fixed C and Python functions spanning common CWEs. The security-cleared rate is the fraction of held-out items ($n{=}16$) for which Semgrep reports the original vulnerability removed in the generated patch.
\end{description}

For all benchmarks, the symbolic reward and the inference reranker see only the training/public signals; held-out tests and the security scan on the held-out set are reserved for evaluation.

\section{Additional Results}
\label{sec:appendix_addl}

\paragraph{Scaling of inference-time selection.}
Table~\ref{tab:appendix_scaling} reports how the four decoding strategies behave as the number of evaluated CodeFlaws bugs grows (DeepSeek-1.3B, $K{=}16$). The ordering is stable from early on: symbolic best-of-$K$ consistently exceeds greedy, random pick consistently trails it, and the oracle marks the achievable headroom. This supports the main-text selector analysis by showing that the result is not an artifact of a particular subset size.

\begin{table}[t]
    \centering
    \small
    \caption{Scaling of inference-time selection on CodeFlaws (DeepSeek-1.3B, $K{=}16$). Values are functional pass@1 (\%) over the first $n$ evaluated bugs, comparing greedy, random, symbolic best-of-$K$, and oracle selection.}
    \setlength{\tabcolsep}{4pt}
    \begin{tabular}{r ccccc}
        \toprule
        $n$ & SFT & SF & Random & Best-of-$K$ & Oracle \\
            & greedy & greedy & pick & (ours) & \\
        \midrule
         75 & 14.7 & 17.3 & 14.7 & 20.0 & 26.7 \\
        150 & 14.0 & 18.0 & 14.0 & 22.0 & 26.0 \\
        225 & 12.9 & 15.1 & 12.0 & 18.2 & 23.1 \\
        300 & 12.3 & 14.0 & 10.7 & 17.7 & 22.7 \\
        375 & 11.7 & 13.1 & 10.4 & 17.1 & 21.9 \\
        \bottomrule
    \end{tabular}
    \label{tab:appendix_scaling}
\end{table}

\paragraph{Surface-similarity metrics saturate.}
Table~\ref{tab:rq1_comparison} reports greedy-decoding CodeBLEU for the SFT-only, RFT-only, and \TheName{} models (all trained from the same base model with the budget-matched schedule). The $\Delta$ column shows that CodeBLEU differences are small and inconsistent across benchmarks: the metric cannot separate the methods even though their deployable repair quality differs sharply (Table~\ref{tab:main_results}). This is the empirical basis for evaluating on execution and security rather than token-overlap metrics.

\begin{table}[tb]
    \centering
    \caption{Greedy-decoding CodeBLEU (\%) for SFT, RFT, and \TheName{}. $\Delta$ is \TheName{}$-$SFT, showing that overlap metrics change only weakly even when deployable repair outcomes improve; best values are in \textbf{bold}.}
    \scriptsize
    \setlength{\tabcolsep}{2.2pt}
    \resizebox{\columnwidth}{!}{%
        \begin{tabular}{l l r r r r}
            \toprule
            Model & Bench. & SFT & RFT & \TheName{} & $\Delta$ \\
            \midrule
            DeepSeek & pyrepair & 77.58 & \textbf{77.78} & 77.53 & $-0.05$ \\
            DeepSeek & SVEN & \textbf{43.88} & 43.81 & 43.73 & $-0.15$ \\
            DeepSeek & CodeFlaws & \textbf{57.90} & 57.25 & 57.87 & $-0.03$ \\
            Llama-3.2 & pyrepair & 76.28 & 76.19 & \textbf{77.30} & $+1.02$ \\
            Llama-3.2 & SVEN & 53.98 & 53.52 & \textbf{54.52} & $+0.54$ \\
            Llama-3.2 & CodeFlaws & 70.38 & 70.43 & \textbf{70.45} & $+0.07$ \\
            Qwen3 & pyrepair & \textbf{82.11} & 80.71 & 80.72 & $-1.39$ \\
            Qwen3 & SVEN & 55.16 & 55.09 & \textbf{55.25} & $+0.09$ \\
            Qwen3 & CodeFlaws & 70.07 & 70.03 & \textbf{70.25} & $+0.18$ \\
            CodeLLaMA & pyrepair & 78.96 & 76.86 & \textbf{79.71} & $+0.75$ \\
            CodeLLaMA & SVEN & 43.18 & 43.26 & \textbf{43.93} & $+0.75$ \\
            CodeLLaMA & CodeFlaws & 57.90 & \textbf{57.92} & 57.74 & $-0.16$ \\
            StarCoder2 & pyrepair & 74.21 & 76.53 & \textbf{80.67} & $+6.46$ \\
            StarCoder2 & SVEN & \textbf{48.63} & 48.17 & 48.22 & $-0.41$ \\
            StarCoder2 & CodeFlaws & \textbf{62.13} & 62.13 & 62.06 & $-0.07$ \\
            \bottomrule
        \end{tabular}%
    }
    \label{tab:rq1_comparison}
\end{table}

\paragraph{Functionality groups (auxiliary diagnostic).}
Table~\ref{tab:functionality_summary} lists the groups used by the auxiliary FixJS/CodeFlaws exact-match analysis in Figure~\ref{fig:functionality_analysis}. This diagnostic is retained only for qualitative context; the paper's primary claims rely on held-out execution and security metrics in Tables~\ref{tab:main_results}--\ref{tab:rq5_ablation_single}.

\begin{table}[tb]
\centering
\caption{Functionality groups used for the auxiliary FixJS/CodeFlaws exact-match analysis in Figure~\ref{fig:functionality_analysis}.}
\label{tab:functionality_summary}
\footnotesize
\setlength{\tabcolsep}{3pt}
\begin{tabular}{l p{0.62\columnwidth}}
\toprule
Dataset & Groups \\
\midrule
JavaScript repair & Network/API, data/object manipulation, DOM/event handling, async callbacks \\
C repair & Math, memory/system code, string processing, I/O and data parsing \\
\bottomrule
\end{tabular}
\end{table}

These appendix diagnostics are intentionally scoped. The scaling table checks selector stability across CodeFlaws subsets, the CodeBLEU table explains why overlap metrics are secondary, and the functionality grouping documents the older exact-match analysis retained for qualitative comparison.

\section{Router Model Training Details}
\label{sec:appendix_4}

\paragraph{Architecture and features.} The Router Model is a lightweight MLP over the normalized compiler-derived feature vector $\tilde{\mathbf{f}}_B$, using two hidden layers of $64$ ReLU units and a sigmoid output for the probability of selecting RFT. In our experiments, $\tilde{\mathbf{f}}_B$ summarizes per-batch AST complexity, CFG depth, and code length. The router is trained concurrently with the Repair Agent using $\mathcal{L}_{\text{Router}}=-\log(p)R_{\text{feedback}}$, where $R_{\text{feedback}}$ is the current batch-loss improvement over a moving-average baseline after normalizing SFT and RFT losses to the same scale.

\section{Future Directions}
\label{sec:appendix_future}

\paragraph{Next steps.} Future versions could add fuzzing, symbolic execution, differential testing, and trace-level assertions, then extend evidence-guided selection to larger repositories with retrieval, build validation, and cross-file reranking.

\paragraph{Selector calibration.} The current selector uses a fixed symbolic scoring rule with a greedy floor. A useful extension is to calibrate the selector across languages and benchmark families, so that evidence sources such as public tests, static analyzers, and structural similarity receive different weights when their reliability changes. This would be especially important for mixed settings where public tests are sparse but static evidence is strong.

\paragraph{Router analysis.} The router is intentionally lightweight, but its decisions provide a useful diagnostic of repair difficulty. Future work could analyze which feature patterns trigger RFT, whether those patterns correspond to semantic bug classes, and how routing changes as stronger base models reduce the need for reward-driven updates.

\paragraph{Artifact extensions.} The released artifact can also be extended with additional candidate traces and selector-feature dumps. Those logs would make it easier to study false positives from the reranker, cases where the greedy floor prevents regressions, and examples where the oracle candidate exists but symbolic ranking fails to select it.

\paragraph{Benchmark coverage.} Finally, broader security benchmarks would help distinguish vulnerabilities that can be repaired through local sanitization from those requiring data-flow or protocol-level reasoning. This would make the router and selector analyses more diagnostic across defect families.

\end{document}